# STUDIES OF THE SUPERCONDUCTING TRAVELLING WAVE CAVITY FOR HIGH GRADIENT LINAC*

Pavel Avrakhov#, Roman Kostin, Alexei Kanareykin (Euclid TechLabs, LLC, Solon, Ohio), Nikolay Solyak, Vyacheslav P. Yakovlev (Fermilab, Batavia)

*Abstract*

Use of a travelling wave (TW) accelerating structure with a small phase advance per cell instead of standing wave may provide a significant increase of accelerating gradient in a superconducting linear accelerator. The TW section achieves an accelerating gradient 1.2-1.4 larger than TESLA-shaped standing wave cavities for the same surface electric and magnetic fields [1]. Recent tests of an L-band single-cell cavity with a waveguide feedback [2] demonstrated an accelerating gradient comparable to the gradient in a single-cell ILC-type cavity from the same manufacturer. This article presents the next stage of the 3-cell TW resonance ring development which will be tested in the travelling wave regime. The main simulation results of the microphonics and Lorentz Force Detuning (LFD) are also considered.

## INTRODUCTION

The primary advantage of a superconducting travelling wave accelerating (STWA) structure is an increased accelerating gradient up to a factor 1.24 for the same surface magnetic and electric field ratios $E_{peak}/E_{acc}$ and $B_{peak}/E_{acc}$ as for re-entrant [3] standing wave (SW) ILC [4] cavity. However, the TW regime for a high gradient linac requires too high level of RF power flux in the accelerating structure. Typical pulse power for L-band TW linac is a few hundred MW which is $10^3$ times more than feeding power for superconducting SW cavities. Concept of the traveling wave resonator with a superconducting TW structure allows to keep pulse power on a reasonable low level

Figure 1 presents the first developed and tested 1.3 GHz prototype of a travelling wave SC resonator which consists of a single-cell cavity and a waveguide (WG) feedback [2]. The 1-cell model has the same shape as a regular full-sized cavity, which helped to understand the problems with mechanical manufacturing, assembly, and welding of this geometry as well as the surface processing issues.

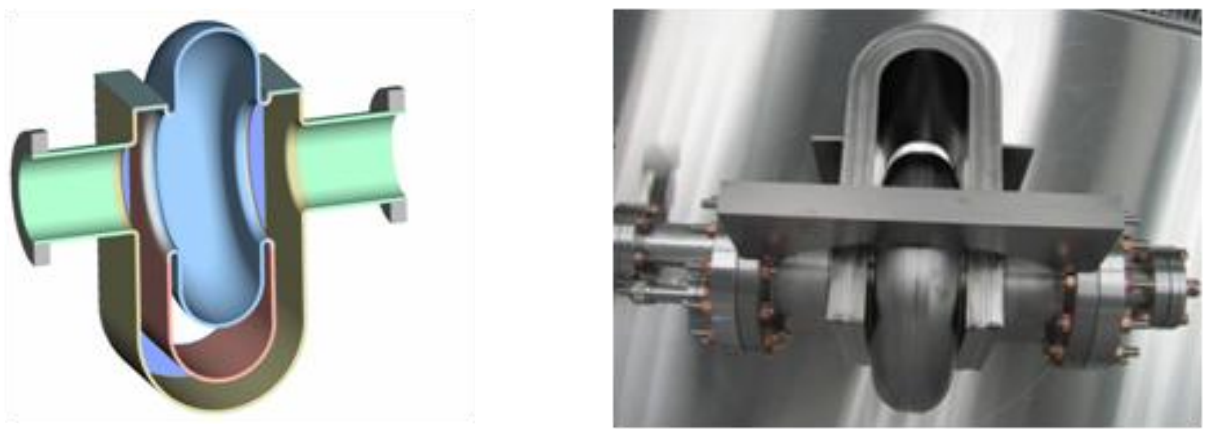

Figure 1: The single-cell model of a travelling wave cavity with waveguide loop.

*Work supported by US Department of Energy

# p.avrakhov@euclidtechlabs.com

Two single cell resonators were manufactured at Advanced Energy System, Inc. (AES) and processed in Argonne and Fermi National Labs. Despite the fact that the STWR processing was without electropolishing, high gradient tests showed that the surface electric and magnetic fields reached the same values as 31 MV/m of TESLA-shaped cavity with no field emission. There were problems with the waveguide processing such as high pressure rinsing. But these problems did not lead to quenching in the waveguide because the surface electric and magnetic field amplitudes in the feedback loop were 2-3 times less than in the accelerating cavity. It should be noted the 1-cell STWR model could be tested only in SW regime.

The next step in SC TW cavity studies will be developing and manufacturing a multicell cavity and testing it in the travelling wave regime.

## A NEW 3-CELL SC TW RESONATOR DESIGN STUDIES

According to plan, a new 3-cell TW cavity design (schematically depicted in Figure 2) was developed [5] and now is prepared for fabrication. In spite of the fact that the 3-cell structure has only one regular cell (in the middle of the cavity), it retains the same field distribution as for the full-sized 1 meter, 15-cell pattern.

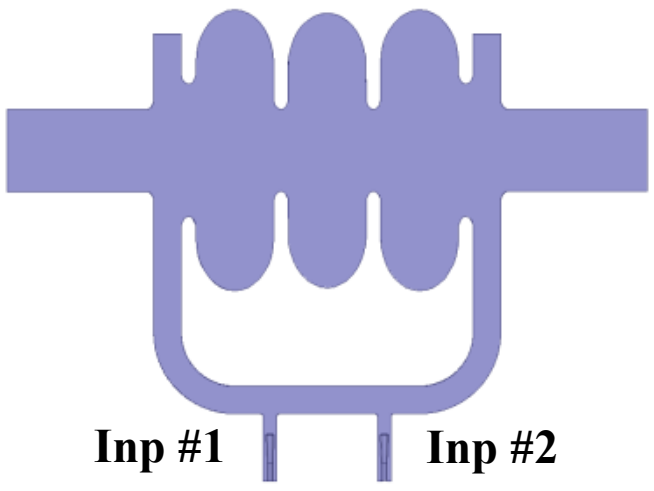

Figure 2: Traveling wave resonator with 3-cell TW accelerating structure and two coaxial couplers.

As a standard for SC accelerating structure research it is testing on vertical test stand (VTS). Low level RF source of about 200-300 W at VTS requires the use of superconducting cavities with a high loaded quality factor $Q_{load}$ of about $10^8$. Well known problems for the SC high $Q_{load}$ resonators are microphonics and Lorentz force detuning at the high gradient testing. But the TW structure detuning has specific feature when compared with the SW cavity - the degradation of the travelling wave regime. Unfortunately, TW operation cannot be restored only by frequency tuning as in the superconducting SW structure. Figure 3 illustrates forward and backward waves in the 3-cell superconducting travelling wave resonator (STWR) with 2.8 mm niobium walls deformed by external pressure. It is evident that a ±0.1 mbar pressure jump completely ruins the travelling wave regime in the non-reinforced STWR.

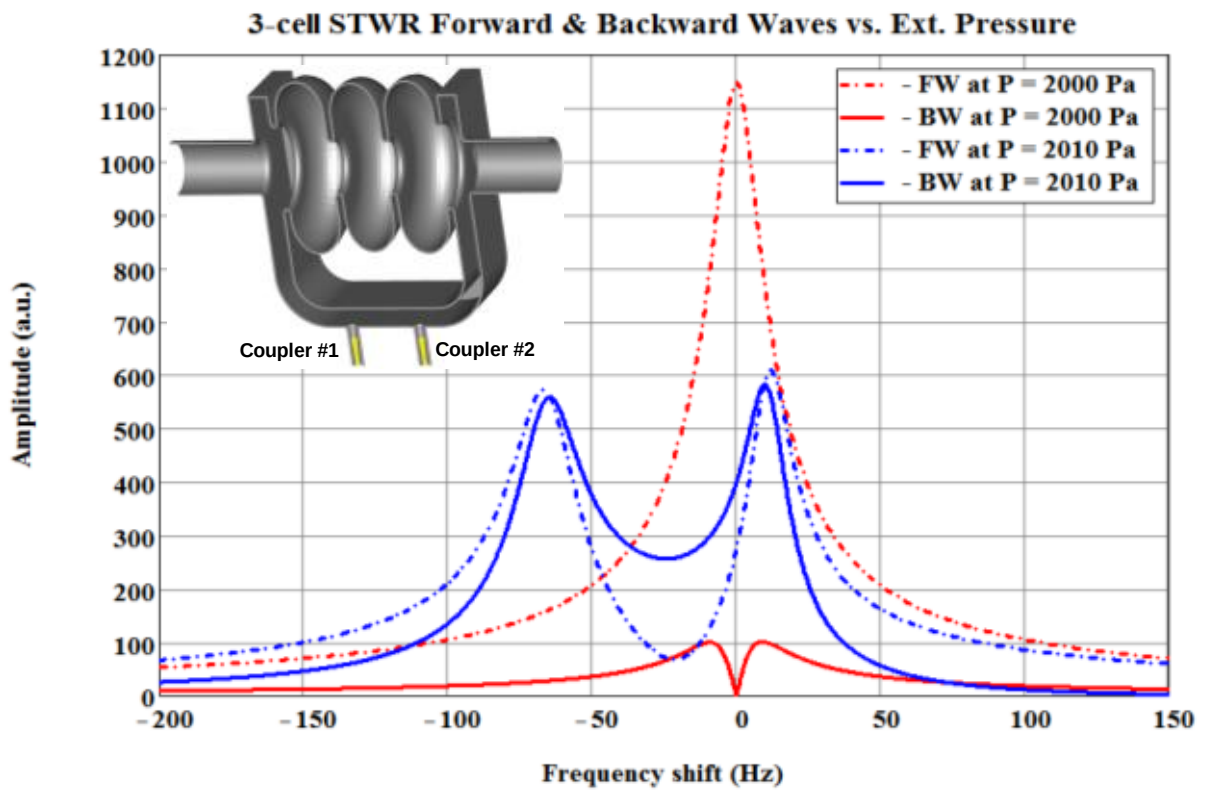

Figure 3: Microphonics detuning of the perfectly matched 3-cell superconducting travelling wave resonator with 2.8 mm niobium walls (pressure jump 10 Pa from 20 mbar to 20.1 mbar).

There are many ways to reinforce a low profile 160mm × 20mm rectangular WG, the most pressure-sensitive part of the STWR, but those involve other technical requirements for SC TW ring like the thermal conductivity of the WG wall and the capability for precise tuning. It imposes restrictions on the wall thickness and the WG loop shape. For instance, a WG loop with walls, more than 4 mm thick, could be overheated in CW regime. The most appropriate design, in our view, it is a waveguide with stiffening ribs on wide, or on both wide and narrow sides. Niobium ribs are a well-known way to reinforce large superconducting cavities such as 325 MHz SSR1, or low frequency 650 MHz 5-cell cavity for Project-X [6].

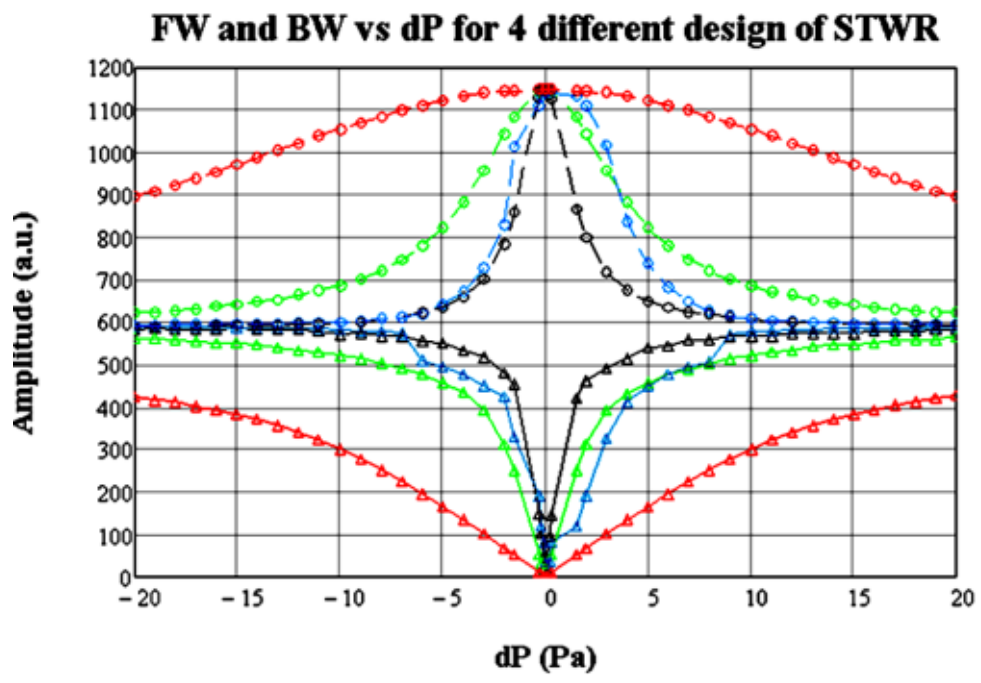

Figure 4: Amplitudes of forward (dash) and backward (solid) waves vs external pressure for 4 different designs of the 3-cell STWR: black curves - the non-reinforced model (all walls 2.8mm); blue curves – the model with cavity ribs only (thickness 4mm); green curves – the model with waveguide ribs only (4mm×10mm and 20mm between ribs); red curves - the most stiffened model with the cavity and waveguide ribs.

Figure 4 presents our efforts to reinforce the 3-cell STWR model to reduce the influence of microphonics. Stiffening ribs on the WG loop and on the cavity part dramatically improves the TW regime. Measured through the critical parameter $d\varphi/dP$, where $d\varphi$ is the phase shift caused by a pressure deviation $dP$, the last design has as a 57 and 12.6 times reduction (compared with the base design) of the microphonics sensitivity for waveguide loop and cavity part respectively.

It should be noted that the presented above results were obtained for fixed amplitudes and phases of the input signals: $P_{inp1} = P_{inp2}$, and a 90° phase difference between inputs. However, these parameters are fully controllable, and can considerably expand the acceptable external pressure variation. Figure 5 shows an example with suppression of microphonics detuning in the most stiffened STWR model only by adjustments of the inputs amplitude and phase. In this case the backward wave at a new operation frequency was completely suppressed, but the forward wave amplitude was kept almost the same (~94%). A 12% increase in power is required to obtain the nominal value of the forward wave amplitude.

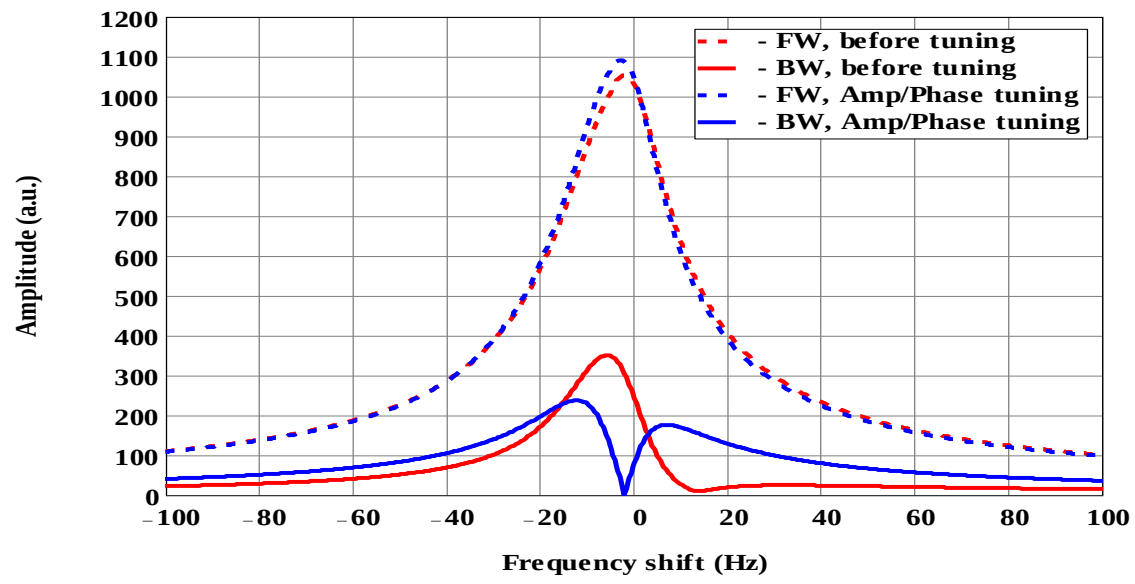

Figure 5: Forward and backward waves (FW & BW) under a 10 Pa external pressure variation in the TW resonator with stiffening ribs on the WG and the cavity parts: red curves – no tuning $P_{inp1}/P_{inp2}$ = 1/1, $\Delta\varphi_{between_inp1\&inp2}$ = 90°; blue curves – after tuning $P_{inp1}/P_{inp2}$ = 1.52/0.48, $\Delta\varphi_{between_inp1\&inp2}$ = 96.3°

High gradient tests of the 1-cell STWR model have shown strong Lorentz forces in SW mode. Figure 6 demonstrates a comparison between the experimental results and the ANSYS simulation of the LFD in the real 1-cell cavity TW1AES002 (see Figure 1). 3-cell model simulations give similar cavity detuning behaviour.

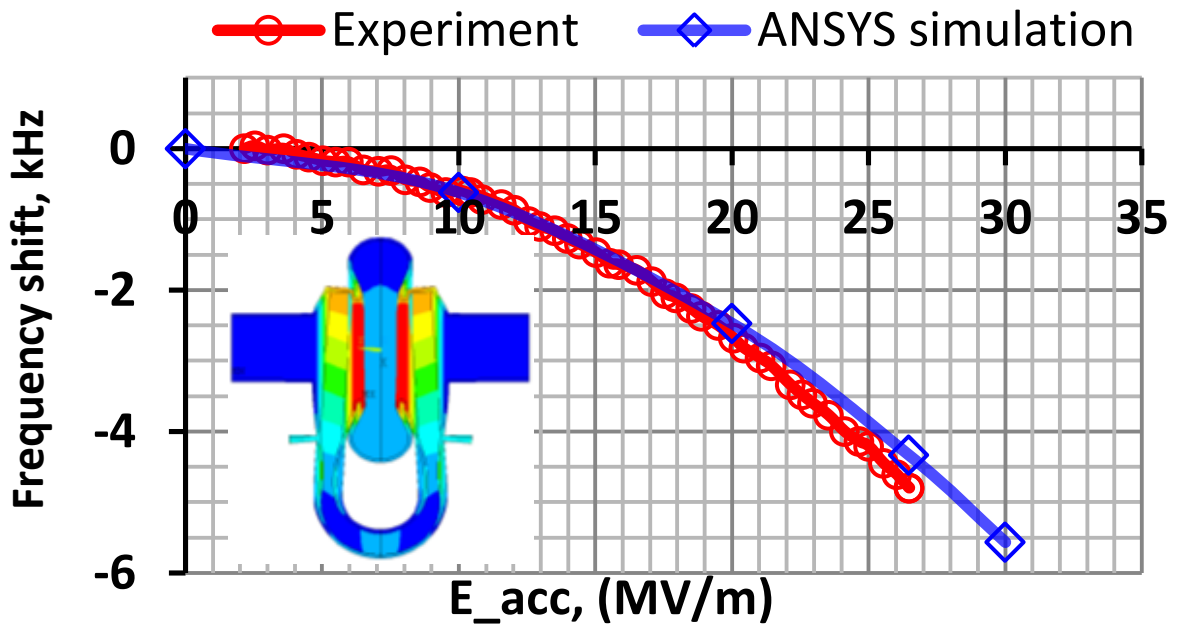

Figure 6: Lorentz force detuning data for 1-cell STWR cavity TW1AES002: experiment and simulation.

The next series of graphs in Figure 7 illustrate simulations of a fine tuning procedure under strong LFD ($E_{acc}$ = 31 MV/m, ring circulating power 631 MW) for the stiffened 3-cell STWR design. The frequency shift for this 3-cell model was -774 Hz and the TW regime was completely destroyed by LFD (Fig. 7-b).

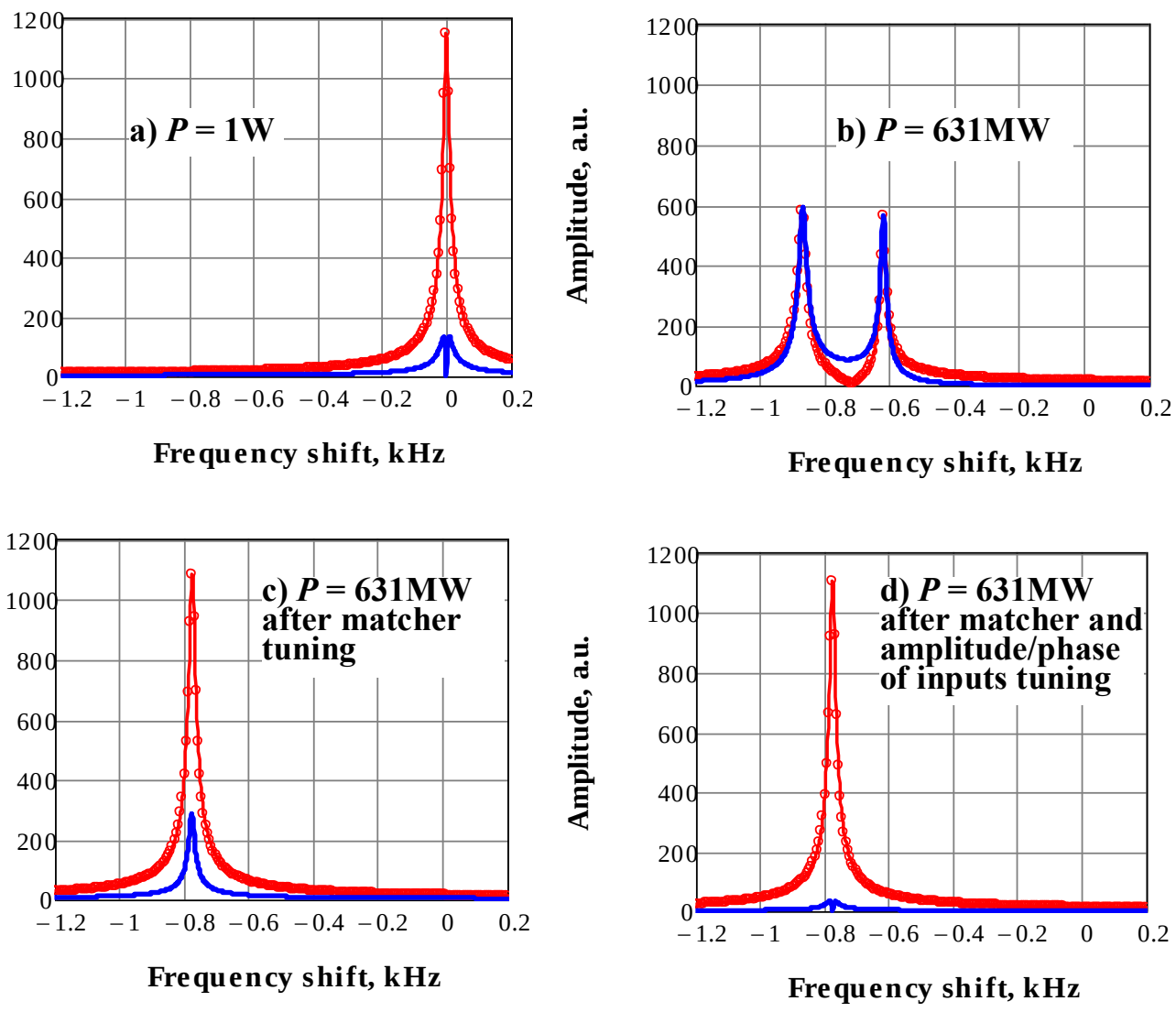

Figure 7: Tuning of the stiffened 3-cell STWR model after strong LFD: a) perfectly matched STWR before LFD; b) strong LFD under circulating power 631 MW; c) tuning by waveguide matcher deformation (+0.61 µm); d) the next stage of tuning is two inputs signal adjusting ($P_{inp1}/P_{inp2} \approx 1.245/0.755$, $\Delta\varphi_{between_inp1\&inp2}$ = 115°, TW mode frequency shift is $\Delta f_{TW}$ = -774.4 Hz)

This example of a strong LFD (about thirty STWR bandwidth) shows two needed stages of tuning – walls shape deformation by a special device named "matcher" (see in Figure 8) and more precise tuning by the redistribution of feeding power in the RF inputs. The first stage is stronger and might adjust both the TW mode and the STWR frequency. The input amplitude and phase control cannot change the cavity frequency and has a narrower acceptable range for detuned parameters. But according to our simulations (see Fig. 5) the fine tuning of RF inputs signals could be enough to withstand microphonics detuning in the stiffened STWR.

## NEW 3-CELL DESIGN OF THE SUPERCONDUCTING TW CAVITY

The present design of the 3-cell STWR based on the obtained simulation results matched with the Fermilab VTS parameters. As SW superconducting accelerating structures the TW cavity has stiffening rings on accelerating cells and, in addition, stiffening ribs on the WG loop (see Figure 8). A new element for a VTS tested prototype will be waveguide matcher which will be used to maintain TW mode in the STWR. This will be achieved through one lever construction with two stepper motors and a movable pushing element. The first stepper motor (not seen in Fig. 8) is for WG shape deformations, which push or pull the pushing element. The second stepper motor is for moving the pushing element along the feedback waveguide in a specially designed cage. These stepper motors can provide up to a 10 nm displacement because of their gear ratios of up to $10^5$ and their 1 mm shaft thread. Necessary matcher requirements obtained through the 3-cell model simulation are range and minimal step for shape deformation ±90µm / 20nm, and the longitudinal position ±4mm / 0.5µm, respectively.

There are five sleeves with conflate flanges for mounting pick-up antennas: three on the WG bend for calibration and TW directivity measurements, and two on the WG loop bottom side for power feeding.

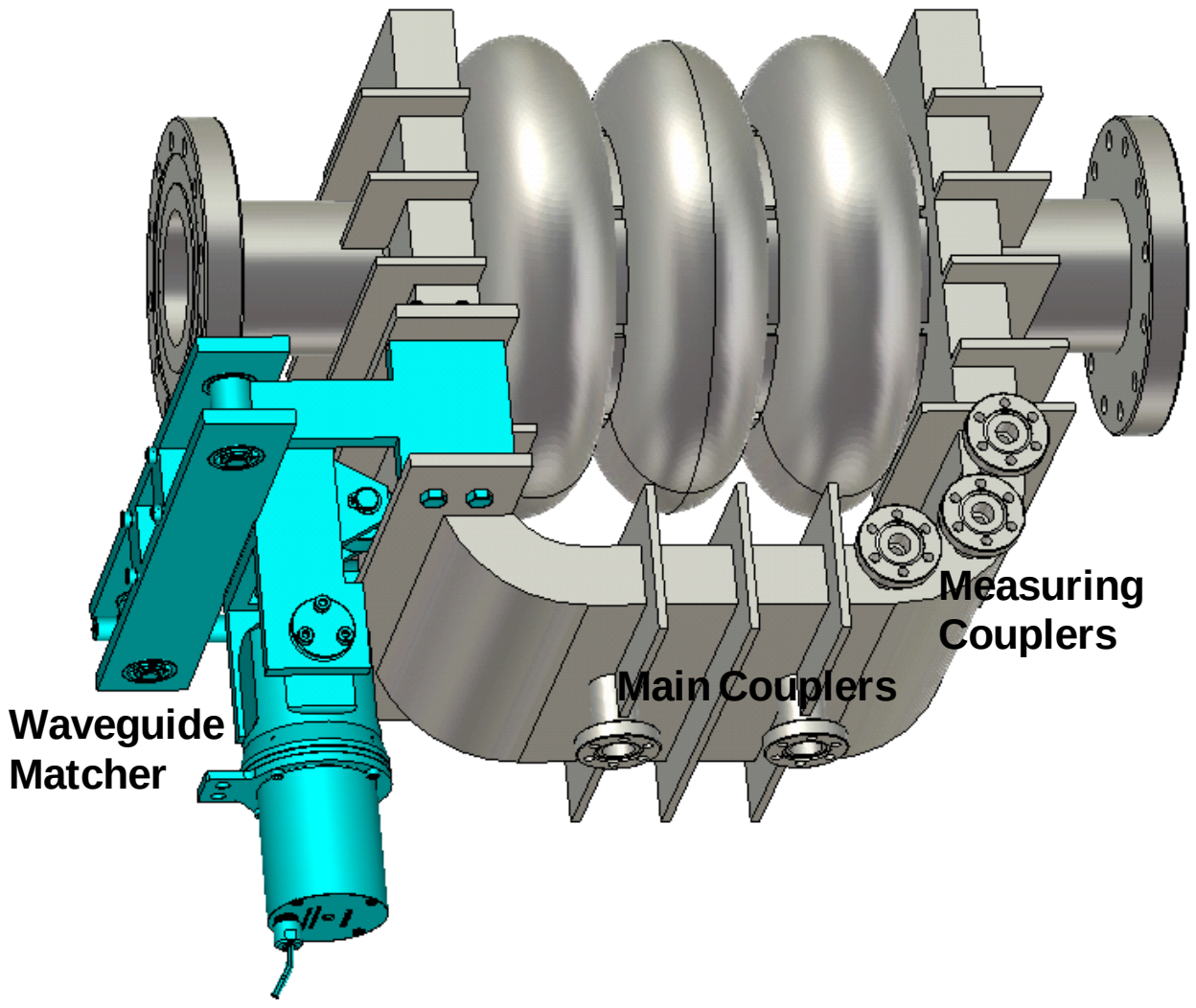

Figure 8: The 3-cell STWR assembly.

## CONCLUSION

The results obtained in the STWR model simulation gave an understanding of the way to build a new 3-cell TW cavity design. Power limitation, strong microphonics, and Lorentz force detuning lead to reinforced niobium cavity, and a new tuning device in the new 3-cell design. Methods to create and support the travelling wave regime in this structure were discussed.